\documentclass[letterpaper, 10pt, conference]{ieeeconf}    

\usepackage[T1]{fontenc}
\usepackage[utf8]{inputenc}
\usepackage{authblk}
\usepackage{float}

\usepackage{amsthm}
\usepackage{amssymb}
\usepackage{amsmath,color}
\usepackage{graphicx}
\usepackage{subcaption}
\usepackage{stmaryrd}
\usepackage{mathbbol}
\usepackage{enumerate}
\usepackage{mathtools}
\usepackage{relsize}
\usepackage{paralist}
\usepackage{bm}
\usepackage{cite,float,algorithmic}
\usepackage[dvipsnames]{xcolor}
\usepackage{tikz}
\usepackage[normalem]{ulem} 
\usepackage{soul}
\usepackage{url}
\let\labelindent\relax
\usepackage{enumerate}
\usepackage[inline]{enumitem}

\newcommand*\MyScale{1}

\tikzset{every picture/.style={scale=\MyScale,}}

\usepackage{pgfplots}
\pgfplotsset{compat=newest}
\pgfplotsset{plot coordinates/math parser=false}

\IEEEoverridecommandlockouts                              
\theoremstyle{plain}

\theoremstyle{plain}
\newtheorem{lem}{Lemma}
\theoremstyle{plain}
\newtheorem{prop}{Proposition}
\theoremstyle{remark}
\newtheorem{rem}{Remark}
\theoremstyle{plain}

\theoremstyle{plain}

\theoremstyle{definition}

\theoremstyle{remark}

\theoremstyle{definition}
\newtheorem{ass}{Assumption}
\theoremstyle{remark}

\floatstyle{ruled}
\newfloat{algorithm}{tbp}{loa}
\providecommand{\algorithmname}{Algorithm}
\floatname{algorithm}{\protect\algorithmname}

\newlength\fheight 
\newlength\fwidth  
\newcommand{\R}{\mathbb{R}}

\newcommand{\E}{\mathcal{E}}
\newcommand{\V}{\mathcal{V}}
\newcommand{\W}{\mathcal{W}}

\DeclareMathOperator*{\argmin}{arg\,min}

\newcommand\scalemath[2]{\scalebox{#1}{\mbox{\ensuremath{\displaystyle #2}}}}

\title{Simultaneous Lane-Keeping and Obstacle Avoidance by Combining Model Predictive Control and Control Barrier Functions}

\author{Sven Br{\"u}ggemann$^*$, Drew Steeves$^*$, and Miroslav Krstic
\thanks{$^*$The first and second authors contributed equally to this work.\\All authors are with the Mechanical \&\ Aerospace Engineering Department, University of California, San Diego, CA 92093-0411, USA, E-mails: {\tt \{sbruegge, dsteeves, krstic\}@eng.ucsd.edu}.}}

\begin{document}

\maketitle

\begin{abstract}
In this work, we combine {Model Predictive Control} (MPC) and Control Barrier Function (CBF) design {methods} to create a hierarchical control law for simultaneous lane-keeping (LK) and obstacle avoidance (OA): at the low level, MPC performs LK via trajectory tracking during nominal operation; and at the high level, different CBF-based safety filters that ensure both LK and OA are designed and compared across some practical scenarios. In particular, we show that Exponential Safety (ESf) and Prescribed-Time Safety (PTSf) filters, which override the MPC control when necessary, result in feasible Quadratic Programs when safety is prioritized appropriately. We additionally investigate control designs subject to input constraints by using Input-Constrained-CBFs. Finally, we compare the performance of combinations of ESf, PTSf, and their input-constrained counterparts with respect to the LK and OA goals in two simulation studies for early- and late-detected obstacle scenarios.
\end{abstract}

\section{Introduction}
Achieving stability while satisfying state and input constraints makes MPC~\cite{rawlings2009model} tremendously appealing to areas such as autonomous systems~\cite{kazban2019,zhangMPC2016,falconeMPC2007,rosaliaAcc2017}. For
example, autonomous vehicles tasked with tracking a
desired trajectory encounter
steering angle input bounds, often employ MPC for motion-planning or trajectory-tracking. However, the constrained optimization problem that forms the backbone of MPC
can have prohibitively long computing times. This is problematic for fast-moving safety-critical systems 
since they demand extremely fast system responses when faced with dangerous system ``perturbations''. Additionally, such systems require state constraints that are time-varying and nonconvex, which also risks recursive feasibility.  

CBF-based \emph{safety filters}
for safety-critical systems have become a popular contender of MPC.
%
Since safety is directly encoded within the CBF, this design method can treat more intricate state constraints while retaining some theoretical guarantees, and prioritizing safety can readily be designed into the safety filter.
In regard to the vehicle \emph{lane-keeping} (LK), MPC is advantageous since, under practical highway-driving assumptions, the (local) vehicle dynamics can be modeled as LTI systems and maximum steering angle inputs are easily handled in the resulting quadratic program (QP). For vehicle \emph{obstacle avoidance} (OA), where (global) nonlinear vehicle dynamics must be considered, CBF-based controllers are desirable since they are computationally cheap, can be designed independent of sampling time, and can natively prioritize safety ``violations''.

In this work, we combine these two control design methods to synthesize a two-layer control law which simultaneously performs vehicle LK and OA for highway driving scenarios.
We investigate high-performance CBFs
which guarantee Prescribed-Time Safety
by using time-varying backstepping to impose safety only over the finite time for which the obstacle is ahead of the vehicle. For simultaneous LK and OA while constraining the steering angle to practical values, we also investigate the marriage between MPC, where input constraints are natively considered, and ICCBFs, where they are integrated within the definition of safety. Thanks to our hierarchical structure, manual LK by a human driver can also be handled while providing safety through automatic OA by our CBF-based design only.
\subsection*{Related work}
The predominant method to design safety filters has been to use a QP which selects the control input ``closest'' (in least-squares sense) to a nominal one, subject to linear inequality constraints that enforce safety.
The authors of~\cite{nguyen2016exponential},
revealed the connection between such Exponential Safety filter (ESf) designs and backstepping through~\cite{krstic2006nonovershooting}. More recently, the time-varying backstepping methods from~\cite{steeves2019prescribed,steeves2022prescribed} were applied in the context of CBF-based safe control design in~\cite{abel2022prescribed} to generate Prescribed-Time Safety (PTSf), which only invokes safety for as long as it is required. In other words, PTSf maximizes the safe operating envelope of systems to retain high nominal performance (by not requiring the system to be ``too safe''); see~\cite{abel2022prescribed} for a complete description of PTSf.

 With the recent Spring of CBFs came their application to safety-critical autonomous systems. For instance, \cite{8114339} proposes to design CBFs offline through a combination of sum-of-squares program and modeling from \cite{rajamani2011vehicle}, and use the related CBF-QP formulation for simultaneous LK and cruise control. In \cite{covorsi}, LK is combined with OA by using mixed-integer formulations for Boolean compositions of multiple discrete-time CBFs, but CBF-QP feasibility is not studied.


{Combining MPC and CBFs has recently gained traction in the research community.
Zhe \emph{et al.} \cite{8431468} design a CBF-CLF hard constraint for the MPC with proven recursive feasibility, producing stability with safety but without computational/feasibility considerations.
Using a multi-rate formulation, in \cite{9442832} a MPC planner is designed in conjunction with a CBF-QP formulation under the assumption that an ICCBF exists. For OA, the authors in \cite{9029446} propose a continuous-time CBF as an additional state constraint for a discrete-time nonlinear MPC. 
The work~\cite{9483029} discusses how discrete-time CBF constraints jeopardize recursive feasibility. To attempt to tackle this issue, the MPC design in~\cite{zeng2021enhancing} is altered by additionally optimizing over the decay rate of the CBF by using slack variables.
}

\section{Lateral vehicle dynamics}
\begin{figure}[t]
    \centering
    \resizebox{.7\columnwidth}{!}{
    \input{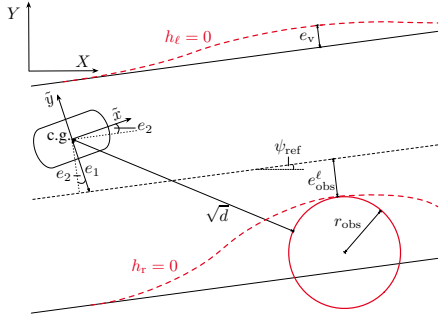}}
  \caption{Obstacle (red circle) on the RHS of the road; lateral distance from center of gravity (c.g.) to the centerline along the vehicle axis, $e_1$, and heading error, $e_2$; here both $e_1,e_2>0$. The shortest distance from the c.g. to the centerline is $e_1\cos(e_2)$. The red dashed lines show the boundaries of the safe set, with tuning parameter $e_{\rm v}$ for possible lane extension. The global coordinates are $X, Y$; the local ones are $\tilde x,\tilde y$.}
    \label{fig:schematic}
\end{figure}

With our underlying goal of LK via trajectory tracking for vehicles during highway-driving conditions, we rely on a dynamic model that governs the local vehicle position, heading and associated velocities (all with respect to the road, see Figure~\ref{fig:schematic}).
We make the following assumption:
\begin{ass}\label{ass:const_long_vel}
The longitudinal vehicle vel.,
$v_{\rm l}$, is constant.
\end{ass}
With Assumption~\ref{ass:const_long_vel}, we obtain the following LTI model that governs the lateral vehicle dynamics~\cite{rajamani2011vehicle}:
%
%
\begin{align}\label{eq:ct_sys}
    \dot x(t)=Ax(t)+Bu(t)+G\dot \psi_{\rm ref}(t),
\end{align}
where $x(t)=\begin{bmatrix}
e_1 & \dot e_1& e_2 & \dot e_2
\end{bmatrix}^\top$, where
$u(t)$ is the front wheel steering angle whose magnitude must not exceed some $u_{\rm max}>0$, and $\dot \psi_{\rm ref}(t)\in\R$ is the (desired) reference yaw rate for $t\in\R_{\geq 0}$.
We denote the components of $A$ and $B$ by $A=(a_{ij})_{1\leq i,j\leq 4}$, $B=(b_{i1})_{1\leq i\leq 4}$, where $b_{i1}=0$ for $i=\{1,3\}$, which makes~\eqref{eq:ct_sys} of relative degree two.
%
If the yaw rate $\dot\psi_{\rm ref}\neq 0$, it is impossible to stabilize the longitudinal and angular errors to the origin, since $G$ lies outside span$(B)$ in general. However, for a fixed yaw rate $\dot \psi_{\rm ref}(t_k)$, as given in \cite[Eq. 3.14]{rajamani2011vehicle}, the steady state and control are\footnote{$l_{\rm r}$ ($l_{\rm f}$) is the distance from the center of gravity to the rear (front) tires (resp.), $k_{\rm v}$ the under-steer gradient, and $\alpha_{\rm r}\dot\psi_{\rm ref}(t)$ is the slip angle of the rear tires.} 
%
\begin{align}
    u_{\rm s}(t)&=\dot\psi_{\rm ref}(t)\left(\frac{l_{\rm f}+l_{\rm r}}{v_{\rm l}}+k_{\rm v}v_{\rm l}\right)\doteq \dot\psi_{\rm ref}(t)\bar u_{\rm s},\label{eq:us}\\
    x_{\rm s}(t)&=\dot\psi_{\rm ref}(t)\begin{bmatrix}
    0 & 0&-\frac{l_{\rm r}}{v_{\rm l}}+\alpha_{\rm r}&0
    \end{bmatrix}^\top\doteq \dot\psi_{\rm ref}(t) \bar x_{\rm s}.\label{eq:xs}
\end{align}

Our LK goal is to drive the system to the yaw-rate-dependent steady-state tuple $(x_{\rm s}(t),u_{\rm s}(t))$, which we detail next using an MPC-based control design.

\section{High-level safety filter}\label{sec:safety_filter}

In this section, we assume that a nominal control input performs LK while driving around a known track (cf. Section~\ref{sec:MPC} for our MPC-based LK design).

Our high-level safety filter is designed to perform both of the LK and OA tasks by employing barrier functions which encode these objectives. It is implemented by the common QP-based approach
\begin{equation}
    \begin{aligned}\label{eq:QP_traditional}
   u_{\rm safe}&=\argmin_{w\in\mathbb{R}}|w-u_{\rm nominal}|^2\\
    \mathrm{s.t.} &\quad a_{\rm i}+b_{\rm i}w\geq 0,\\
    &\quad a_{\rm k}+b_{\rm k}w\geq 0,
    \end{aligned}
\end{equation}
where one inequality constraint is used to encode the LK objective, and the other encodes the OA objective. In order to generate safety overriding controllers which can simultaneously satisfy both objectives (and in particular, satisfy the so-called control-sharing property~\cite{xu2018constrained}, which is a necessary and sufficient condition for~\eqref{eq:QP_traditional} to be feasible), we propose a barrier function which smoothly transitions from one of the roadway lane barriers to one encapsulating the obstacle, whereas the other lane barrier either remains the same or expands in a direction away from the obstacle (see Figure~\ref{fig:schematic}).

Note that lane modification is not inherent or immediately available to the MPC design method: in most applications, modifying any existing state constraints in real-time brings into question recursive feasibility. It is the combination of a low-level MPC design and a high-level safety filter design that allows us to prescribe how the vehicle passes the obstacle while retaining theoretical guarantees, and allows us to characterize what violation may be needed with respect to the original lane barriers in order to ensure that~\eqref{eq:QP_traditional} remains feasible.

\subsection{Control Barrier Function Design for LK and OA}\label{sub:CBF_design}
We begin by designing our barrier functions (BFs) that encode vehicle safety w.r.t. LK and OA. We first define the squared relative distance from vehicle to obstacle,
\begin{align*}
    \scalemath{0.88}{d(e_1,e_2):=\left(X_{\rm car}(e_1,e_2)-X_{\rm obs}\right)^2+\left(Y_{\rm car}(e_1,e_2)-Y_{\rm obs}\right)^2-r_{\rm obs}^2,}
\end{align*}
where $r_{\rm obs}$ denotes the radius of the obstacle-encompassing circle, and $\left(X_{\rm car}, Y_{\rm car}\right)$, $\left(X_{\rm obs}, Y_{\rm obs}\right)$ denote the global coordinates of the vehicle and the obstacle, respectively. Next, we define
the smooth step function which depends on the relative distance, $d(e_1,e_2)$, and the parameter $\delta_2>0$ which is the relative detection distance:
\begin{align}\label{eq:Phi_def}
    \Phi(d)&=\begin{cases}
    0,&d\geq \delta_2^2,\\
    \mathrm{exp}\left(1-\frac{\delta_2^2}{\delta_2^2-d}\right),& 0<d<\delta_2^2,\\
    1,&d\leq 0.
    \end{cases}
\end{align}
For $w_{\rm l}>0$ denoting the width of the lane and $e_{\rm{obs}}^{\rm{\ell}}$ denoting the shortest distance of the obstacle-encompassing circle to the centerline (see Figure \ref{fig:schematic}), we define the left BF (ensuring LK or a certain violation of of LK)
\begin{align}\label{eq:h_l_def}
    \scalemath{0.84}{h_{\rm \ell}
    :=\Phi(d)\left(\frac{w_{\rm l}}{2}-e_1\cos\left(e_2\right)+e_{\rm v}\right)+\left(1-\Phi(d)\right)\left(\frac{w_{\rm l}}{2}-e_1\cos\left(e_2\right)\right),}
\end{align}
where $e_{\rm v}\geq 0$ is a design parameter which allows the vehicle to utilize a portion of the highway shoulder or a left-hand lane (assuming no oncoming traffic). By selecting $e_{\rm v}=0$, the left barrier simplifies to $h_{\rm \ell}(e_1,e_2)=\frac{w_{\rm l}}{2}-e_1$, that is, it remains constant regardless of the vehicle's proximity to the obstacle. We define the right BF (ensuring OA) as
\begin{align}\label{eq:h_r_def}
    \scalemath{0.88}{h_{\rm r}:=\Phi(d)\left(e_1\cos\left(e_2\right)-e_{\rm obs}^{\ell}\right)+\left(1-\Phi(d)\right)\left(\frac{w_{\rm l}}{2}+e_1\cos\left(e_2\right)\right).}
\end{align}
While the choice $e_{\rm v}=0$ is made throughout Section~\ref{sec:sim} with success, to establish theoretical safety guarantees (cf. Lemma~\ref{lemma:control_sharing} and Remark~\ref{rk:control_sharing_result}), we must prioritize OA over LK by selecting $e_{\rm v}>0$.

For the purpose of presentation clarity, we assume w.l.o.g. that the (geometric) center of the obstacle-encompassing circle is on the right side of the road centerline, i.e., that $e_{\rm{obs}}^{\rm{\ell}}<0$, and that we seek to pass on the left side of the obstacle. Figure~\ref{fig:schematic} depicts the left and right BFs as red dashed lines.
%
We define the safe sets
\begin{align*}
    \scalemath{0.88}{\mathcal{S}_{\rm \ell}:=\{x\in\mathbb{R}^4\;|\;h_{\rm \ell}(e_1,e_2)\geq 0\},}\;\;\scalemath{0.88}{\mathcal{S}_{\rm r}:=\{x\in\mathbb{R}^4\;|\;h_{\rm r}(e_1,e_2)\geq 0\};}
\end{align*}
our goal is to design MPC-overriding control laws which renders $\mathcal{S}_{\rm l}\cap\mathcal{S}_{\rm r}$ positively-invariant.

While the MPC design for lane-keeping is performed using the linear local dynamics~\eqref{eq:ct_sys}, the safety filter design, which invokes the vehicle's global position in~\eqref{eq:Phi_def},
involves the vehicle's nonlinear dynamics: the vehicle coordinates $\left(X_{\rm car}, Y_{\rm car}\right)$ are related to the rel.-deg.-two system~\eqref{eq:ct_sys} by
\begin{align*}
    \scalemath{0.93}{\frac{dX_{\rm car}}{dt}(e_1,\dot{e}_1,e_2)=v_{\rm{l}}\cos\left(e_2+\psi_{\rm r}\right)-\left(\dot{e}_1-v_{\rm{l}}e_2\right)\sin\left(e_2+\psi_{\rm r}\right),}\\
    \scalemath{0.93}{\frac{dY_{\rm car}}{dt}(e_1,\dot{e}_1,e_2)=v_{\rm{l}}\sin\left(e_2+\psi_{\rm r}\right)+\left(\dot{e}_1-v_{\rm{l}}e_2\right)\cos\left(e_2+\psi_{\rm r}\right).}
\end{align*}
Hence, it follows from the vehicle's (global) velocities and the dynamics that~\eqref{eq:h_l_def},~\eqref{eq:h_r_def} are rel.-deg.-two CBFs.


\subsection{Exponential Safety by (Time-Invariant) Backstepping}\label{sub:ESf}

Next, we introduce the time-invariant backstepping method which leads to explicit characterizations for the safety overriding controllers, referred to as ESf control designs, which render either $\mathcal{S}_{\ell}$ or $\mathcal{S}_{\rm r}$ positively-invariant. This method is a specific case of more general CBF-safety designs, where the inequalities in~\eqref{eq:QP_traditional} are typically constructed using class-$\mathcal{K}$ functions of the CBFs and their derivatives; the backstepping framework presents a direct connection between safety conservatism and performance via gain tuning~(cf. Section~\ref{sub:perf_safety_tradeoff} for details).

For $\rm{i}\in\{\rm{\ell},\rm{r}\}$, we define
\begin{align}
    h_{\rm{i},1}(e_1,e_2)&=h_{\rm i}(e_1,e_2),\\
    h_{\rm{i},2}(x)&=\frac{d}{dt}h_{\rm{i},1}(x)+c_{\rm{i},1}h_{\rm{i},1}(e_1,e_2);\label{eq:hi_backstepping_step}
\end{align}
by selecting the gains to satisfy
\begin{align}\label{eq:ESf_gain_c1}
    c_{\rm{i},1}>\max\left\{0,-\frac{\frac{d}{dt}h_{\rm{i},1}(x(0))}{h_{\rm{i},1}(e_1(0),e_2(0))}\right\},
\end{align}
we ensure that $h_{\rm{i},2}(x(0))>0$. To guarantee $h_{\rm{i},1}(e_1(t),e_2(t))\geq 0$ for all $t\in[0,\infty)$, or in other words, that $\mathcal{S}_{\rm{i}}$ remains positively-invariant indefinitely, we must additionally show that
\begin{align}\label{eq:hi_2_inequality}
    \frac{d}{dt}h_{\rm{i},2}+c_{\rm{i},2}h_{\rm{i},2}\geq 0,
\end{align}
for $c_{\rm{i},2}>0$; indeed, the barrier constraint~\eqref{eq:hi_2_inequality} is exactly~\cite[Equ. 7]{xu2018constrained} for $r=2$, $a_1=c_{\rm{i},1}+c_{\rm{i},2}$, and $a_2=c_{\rm{i},1}c_{\rm{i},2}$, since
\begin{align}
    \scalemath{0.95}{c_{\rm{i},2}h_{\rm{i},2}}\;&\scalemath{0.95}{=c_{\rm{i},2}\left(\frac{d}{dt}h_{\rm{i},1}+c_{\rm{i},1}h_{\rm{i},1}\right)=:c_{\rm{i},2}L_fh_{\rm{i},1}+c_{\rm{i},1}c_{\rm{i},2}h_{\rm{i},1},}\label{eq:c*h_2_calc}\\
    \scalemath{0.95}{\frac{d}{dt}h_{\rm{i},2}}&\scalemath{0.95}{:=L_f^2h_{\rm{i},1}+L_gL_fh_{\rm{i},1}u+c_{\rm{i},1}L_fh_{\rm{i},1}.}\label{eq:ddt_h2_calc}
\end{align}
A safety filter can then be designed as the QP~\eqref{eq:QP_traditional}, with $a_{\rm i}:=L_f^2h_{\rm{i},1}+\left(c_{\rm{i},1}+c_{\rm{i},2} \right)L_fh_{\rm{i},1}+c_{\rm{i},1}c_{\rm{i},2}h_{\rm{i},1}$, $b_{\rm i}:=L_gL_fh_{\rm{i},1}$, $a_{\rm k}=0$, and $b_{\rm k}=0$. Equivalently, we can select
\begin{align}\label{eq:u_override_ESf}
    \scalemath{0.9}{u_{\rm{i}, \rm{override}}=-\frac{L_f^2h_{\rm{i},1}+\left(c_{\rm{i},1}+c_{\rm{i},2}\right)L_fh_{\rm{i},1}+c_{\rm{i},1}c_{\rm{i},2}h_{\rm{i},1}}{L_gL_fh_{\rm{i},1}},}
\end{align}
which is well-defined so long as $L_gL_fh_{\rm{i},1}$ never equals zero, and implement
\begin{equation}
    \begin{aligned}\label{eq:QP_new}
       u_{\rm safe}&=\argmin_{w\in\mathbb{R}}|w-u_{\rm MPC}|^2\\
        \mathrm{s.t.} &\quad w\geq u_{\rm override}.
    \end{aligned}
\end{equation}
In scenarios where $L_gL_fh_{\rm{i},1}=0$, the \emph{validity} of $h_{\rm{i}}$ as a CBF depends on whether~\eqref{eq:hi_2_inequality} holds despite $L_gL_fh_{\rm{i},1}=0$ (that is, $h_{\rm{i}}$ is a \emph{valid} CBF if~\eqref{eq:hi_2_inequality} holds true when $h_{\rm{i}}\geq 0$ and despite the control having no effect on~\eqref{eq:hi_2_inequality}).

In practice, verifying the validity of a candidate CBF is difficult when $L_gL_fh_{\rm{i},1}=0$ at some time instances, since the system states are arbitrary and the sign of the Lie derivative terms in~\eqref{eq:c*h_2_calc},~\eqref{eq:ddt_h2_calc} may be state-dependent and large (as is the case here for the sinusoidal nonlinearities). This is often resolved in practice (e.g.,~\cite{ames2016control,hsu2015control,wu2015safety}) by modifying the barrier constraint in~\eqref{eq:QP_traditional} to be a soft constraint. For the CBFs~\eqref{eq:h_l_def} and~\eqref{eq:h_r_def}, we compute
\begin{align}\label{eq:LgLfhl}
    &\scalemath{0.85}{L_gL_fh_{\rm{\ell},1}=-b_{21}\cos(e_2)+b_{41}e_1\sin(e_2)+\frac{2b_{21}\delta_2^2}{(\delta_2^2-d(e_1))^2}\Phi(d(e_1))e_{\rm v}}\nonumber\\
    &\scalemath{0.85}{\times\Big[\left(X_{\rm car}-X_{\rm obs}\right)\sin(e_2+\psi_{\rm r})-\left(Y_{\rm car}-Y_{\rm obs}\right)\cos(e_2+\psi_{\rm r})\Big],}
\end{align}
and
\begin{align}\label{eq:LgLfhr}
    &\scalemath{0.83}{L_gL_fh_{\rm{r},1}=b_{21}\cos(e_2)-b_{41}e_1\sin(e_2)-\frac{2b_{21}\delta_2^2}{(\delta_2^2-d(e_1))^2}\Phi(d(e_1))}\\
    &\hspace{-1mm}\scalemath{0.83}{\times\hspace{-1mm}\left(\frac{w_l}{2}+e_{\rm obs}^{\ell}\right)\hspace{-1mm}\left[\left(X_{\rm car}-X_{\rm obs}\right)\sin(e_2+\psi_{\rm r})\hspace{-1mm}-\hspace{-1mm}\left(Y_{\rm car}-Y_{\rm obs}\right)\cos(e_2+\psi_{\rm r})\right],}\nonumber
\end{align}
which \emph{may} equate to zero for certain vehicle headings, heading errors, and relative distances between vehicle and obstacle; this is one caveat of using Assumption~\ref{ass:const_long_vel} to obtain~\eqref{eq:ct_sys} (which effectively renders our nonholonomic model as \emph{underactuated}) since otherwise, another control term would appear to prevent $L_gL_fh_{\rm{i},1}=0$. However, in the numerous closed-loop simulation studies performed (some of which are featured in Section~\ref{sec:sim}), $L_gL_fh_{\rm{i},1}=0$ never occured. Note that $\lim_{d(e_1)\rightarrow \delta_2^2}\frac{\Phi(d(e_1))}{(\delta_2^2-d(e_1))^k}=0$ for any $k\in\mathbb{N}$.

By selecting $e_{\rm v}=\left(\frac{w_l}{2}+e_{\rm obs}^{\ell}\right)$ in~\eqref{eq:LgLfhl}, we obtain $L_gL_fh_{\rm{\ell},1}=-L_gL_fh_{\rm{r},1}$, which generates the following result.

\begin{lem}\label{lemma:control_sharing}
Suppose a vehicle governed by~\eqref{eq:ct_sys} detects an obstacle and has access to $d(e_1,e_2)$. Under Assumption~\ref{ass:const_long_vel}, and if $L_gL_fh_{\rm{i},1}(x)\neq0$ for all $x\in\mathcal{S}_{\rm i}$, $\rm{i}\in\{\ell,\rm{r}\}$, if we permit the left-lane expansion
\begin{align}\label{eq:ev_selection}
    e_{\rm v}=\left(\frac{w_l}{2}+e_{\rm obs}^{\ell}\right)>0,\quad\text{where}\quad e_{\rm obs}^{\ell}<0,
\end{align}
then for $\rm{i}\in\{\ell,\rm{r}\}$ and the overriding controller~\eqref{eq:u_override_ESf}, if we select $c_{\mathrm{\ell},j}=c_{\mathrm{r},j}>0$ for $j=1,2$, then
\begin{align}\label{eq:CS_inequality}
    u_{\ell, \rm{override}}-u_{\rm{r}, \rm{override}}=\left(L_gL_fh_{\ell,1} \right)^{-1}c_{\ell,1}c_{\ell,2}w_l.
\end{align}
In other words, the CBFs~\eqref{eq:h_l_def},~\eqref{eq:h_r_def} have the control-sharing property~\cite{xu2018constrained}
, and hence generate the feasible QP
\begin{equation}
    \begin{aligned}\label{eq:QP_CS_lemma}
       u_{\rm safe}&=\argmin_{w\in\mathbb{R}}|w-u_{\rm MPC}|^2\\
        \mathrm{s.t.} &\quad u_{\rm{r}, \rm{override}}\leq w\leq u_{\ell, \rm{override}},
    \end{aligned}
\end{equation}
when $L_gL_fh_{\ell,1}<0$ (otherwise, the barrier inequalities must be flipped). Moreover, if we additionally select $c_{\ell,1}$ to satisfy~\eqref{eq:ESf_gain_c1},
then the safety filter~\eqref{eq:QP_CS_lemma} ensures that $\mathcal{S}_{\ell}\cap\mathcal{S}_{\rm{r}}\neq\emptyset$ remains positively-invariant for~\eqref{eq:ct_sys} for all times provided that $x(0)\in\mathcal{S}_{\ell}\cap\mathcal{S}_{\rm{r}}$, and~\eqref{eq:ct_sys} with~\eqref{eq:QP_CS_lemma} is ESf.
\end{lem}
The proof of Lemma~\ref{lemma:control_sharing} directly follows from computing the overriding controllers, the selection~\eqref{eq:ev_selection}, and both~\cite[Thm. 1]{xu2018constrained},~\cite[Cor. 2]{nguyen2016exponential}.
\begin{rem}\label{rk:control_sharing_result}
While Lemma~\ref{lemma:control_sharing} prioritizes OA over LK by permitting the vehicle to ``violate'' the left lane marker by up to an additional half lane width in~\eqref{eq:ev_selection}, these conditions are encountered in practice: vehicles can often access a left-hand lane (assuming no oncoming vehicles) or a highway shoulder. The result illustrates the conservativeness required to establish theoretical guarantees when using multiple CBFs, since~\eqref{eq:CS_inequality} $e_{\rm v}=0$ was sufficient in our simulation studies.
\end{rem}

\subsection{Balancing Performance and Safety by ESf and PTSf}\label{sub:perf_safety_tradeoff}
We classify ``high performance'' as imposing the minimum amount of control intervention to retain safety. If we select large gains $c_{\mathrm{i},j}$ for $\rm{i}\in\{\ell,\rm{r}\}$, $j=1,2$, then the exponential decay in time of $h_{\mathrm{i},j}$ will be large, causing the safe operating envelope $\mathcal{S}_{\rm{i}}$ to become larger.
In other words, a larger gain makes safety less conservative, causing less intervention by the safety filter with the nominal controller, and allowing the vehicle to perform as intended. One caveat to selecting large constant gains is that the initial behavior of the safety filter~\eqref{eq:QP_CS_lemma} can exhibit a drastic intervention when the nominal control is {first} unsafe, as is exhibited in Figure~\ref{fig:TI_PI_u} and discussed therein; this behavior is not an issue with smaller gains. These large interventions are particularly problematic when actuator constraints are present. To circumvent them while retaining the intended closed-loop performance, we investigate PTSf by using time-varying-gains for~\eqref{eq:ct_sys},~\eqref{eq:QP_CS_lemma} which start small (when $h_{\mathrm{i},j}$ is large) and grow large (as $h_{\mathrm{i},j}$ becomes small, as the vehicle approaches the obstacle). 

\subsection{Prescribed-Time Safety by Time-Varying Backstepping}\label{sub:PTSf}

Since the obstacle threatens the safe operation of the vehicle only while the vehicle is moving towards it, the OA problem is finite-time in nature, and we would like the safety filter to only perform LK after passing the obstacle. One way to accomplish this is to revert from the OA barrier constraint back to one which encodes LK safety; this behavior is in fact exhibited in a smooth fashion by our CBF design~\eqref{eq:h_r_def}. But we would also like to balance safety and performance via gain tuning without producing large overriding control inputs. To this end, we turn to PTSf designs.

PTSf uses time-varying gains to not only balance performance and safety by small overriding inputs, but to enforce safety only for as long as required. PTSf designs retain $h_{\mathrm{i},j}(t)\geq 0$ but drive $h_{\mathrm{i},j}(t)\rightarrow 0$ \emph{within a finite time} that can be a priori prescribed, for $i\in\{\ell,\rm{r}\}$ and $j=1,2$; in contrast, ESf requires $t\rightarrow \infty$ for $h_{\mathrm{i},j}(t)\rightarrow 0$ (see~\cite{abel2022} for a more extensive discussion on PTSf).

PTSf designs rely on the following time-varying blow-up function
\begin{align}
    \scalemath{0.95}{\mu_2(t-t_{\rm{obs}},T):=\frac{1}{\left(1-\frac{t-t_{\rm{obs}}}{T}\right)^2},\quad t\in[t_{\rm obs},t_{\rm obs}+T],}
\end{align}
which equals one at the detection time $t_{\rm{obs}}$ (when $d(e_1,e_2)=\delta_2^2$ in~\eqref{eq:Phi_def}) but equals infinity at the \emph{passing time} $t_{\rm{obs}}+T$.

Since the longitudinal vehicle velocity $v_{\rm{l}}$ is constant, we can estimate the passing time quite accurately for highway driving scenarios, by using the following relation to numerically solve for the $T$:
\begin{align}\label{eq:passing_time_approx}
    \scalemath{0.83}{d_{\rm{obs}}^{\rm{path}}}\;&\scalemath{0.83}{=\Bigg[\left(\int_{t_{\rm{obs}}}^Tv_{\rm{l}}\cos\left(\psi_{\rm{ref}}(t)\right)dt\right)^2+\left(\int_{t_{\rm{obs}}}^Tv_{\rm{l}}\sin\left(\psi_{\rm{ref}}(t)\right)dt\right)^2\Bigg]^{1/2},}
\end{align}
where $d_{\rm{obs}}^{\rm{path}}$ is the relative distance between the end of the obstacle and the vehicle along the path at time $t=t_{\rm{obs}}$.

We now exchange the constant gains in~\eqref{eq:u_override_ESf} with the following time-varying gains: for $t\in[t_{\rm{obs}},t_{\rm{obs}}+T)$, we define
\begin{align}
    c_{\mathrm{i},j}(t)=c_{\mathrm{i},j}^0\mu_2(t-t_{\rm{obs}},T),\quad i\in\{\ell,\mathrm{r}\},\;j=1,2,
\end{align}
where $c_{\mathrm{i},j}^0>0$ also satisfying~\eqref{eq:ESf_gain_c1} are the \emph{initial} gains at detection time and can be chosen to be as small as possible while retaining a large safe operating envelope. Employing these time-varying gains generates the following result, which is a consequence of Lemma~\ref{lemma:control_sharing} and the treatment in the proof of~\cite[Thm. 1]{abel2022}.
\begin{prop}\label{prop:PTSf}
Suppose a vehicle governed by~\eqref{eq:ct_sys} detects an obstacle and has access to $d(e_1,e_2)$. Under Assumption~\ref{ass:const_long_vel}, and if $L_gL_fh_{\rm{i},1}(x)\neq0$ for all $x\in\mathcal{S}_{\rm i}$, $\rm{i}\in\{\ell,\rm{r}\}$, if we select~\eqref{eq:ev_selection}, then for $\rm{i}\in\{\ell,\rm{r}\}$ and
\begin{align}\label{eq:u_override_PTSf}
   \scalemath{0.9}{ u_{\rm{i}, \rm{override}}}\;&\scalemath{0.9}{=-\frac{L_f^2h_{\rm{i},1}+\left(c_{\rm{i},1}+c_{\rm{i},2}\right)L_fh_{\rm{i},1}+\left(\dot{c}_{\rm{i},1}+c_{\rm{i},1}c_{\rm{i},2}\right)h_{\rm{i},1}}{L_gL_fh_{\rm{i},1}},}
\end{align}
if we select $c_{\mathrm{\ell},j}=c_{\mathrm{r},j}>0$ for $j=1,2$, then
\begin{align}\label{eq:CS_inequality_PTSf}
    \scalemath{0.9}{u_{\ell, \rm{override}}-u_{\rm{r}, \rm{override}}=\left(L_gL_fh_{\ell,1} \right)^{-1}\left(\dot{c}_{\ell,1}+c_{\ell,1}c_{\ell,2}\right)w_l.}
\end{align}
In other words, the CBFs~\eqref{eq:h_l_def},~\eqref{eq:h_r_def} have the control-sharing property, and hence generate the feasible QP
\begin{equation}
    \begin{aligned}\label{eq:QP_CS_lemma_PTSf}
       u_{\rm safe}&=\argmin_{w\in\mathbb{R}}|w-u_{\rm MPC}|^2\\
        \mathrm{s.t.} &\quad u_{\rm{r}, \rm{override}}\leq w\leq u_{\ell, \rm{override}},
    \end{aligned}
\end{equation}
when $L_gL_fh_{\ell,1}<0$ (otherwise, the barrier inequalities must be flipped). If we additionally select $c_{\ell,1}$ to satisfy~\eqref{eq:ESf_gain_c1},
then the safety filter~\eqref{eq:QP_CS_lemma_PTSf} ensures that $\mathcal{S}_{\ell}\cap\mathcal{S}_{\rm{r}}\neq\emptyset$ remains positively-invariant for~\eqref{eq:ct_sys} over the interval $[t_{\rm{obs}},t_{\rm{obs}}+T)$ provided that $x(t_{\rm obs})\in\mathcal{S}_{\ell}\cap\mathcal{S}_{\rm{r}}$, and~\eqref{eq:ct_sys} with~\eqref{eq:QP_CS_lemma_PTSf} is PTSf. Moreover, the time-varying overriding control laws are uniformly bounded for $t\in[t_{\rm{obs}},t_{\rm{obs}}+T]$.
\end{prop}
Absent from Proposition~\ref{prop:PTSf} are closed-loop system guarantees after the passing time $t_{\rm{obs}}+T$, which are required since our LK objective has not expired. Handling the PTSf filter behavior after the passing time can be done using a smooth ramp function, as detailed in~\cite[Equs. (10), (11)]{abel2022}, and as implemented in Section~\ref{sec:sim}.

Moreover, the time-varying filter~\eqref{eq:QP_CS_lemma} is not required for the LK objective, since LK persists after the passing time. In Section~\ref{sec:sim}, we use~\eqref{eq:u_override_ESf} for $\mathrm{i}=\ell$ and we use~\eqref{eq:u_override_PTSf} for $\rm i=r$; however, with this combination, the control-sharing property becomes difficult to establish and requires further study which, due to space constraints, will be featured elsewhere.

\subsection{Input-Constrained CBFs (ICCBFs)}\label{sub:ICCBF}

Given our hierarchical approach to solve the LK and OA problems for vehicles travelling on the highway,
one glaring incompatibility between MPC- and CBF-based control designs is that the former can handle input constraints, whereas up until recently, the latter could not.

The work~\cite{agrawal2021safe} introduces ICCBF, whose designs that are similar to CBF designs, except they restrict the safe sets $\mathcal{S}_{\rm i}$ further by iteratively removing states from which system safety can only be achieved by violating the input constraints. In the (time-invariant) backstepping framework, the first iteration of the ICCBF design replaces~\eqref{eq:hi_backstepping_step},~\eqref{eq:hi_2_inequality} by
\begin{align}
    \frac{d}{dt}h_{\rm{i},1}(x)&=-c_{\rm{i},1}h_{\rm{i},1}(e_1,e_2)+h_{\rm{i},2}(x),\\
    \frac{d}{dt}h_{\rm{i},2}&\geq-c_{\rm{i},2}h_{\rm{i},2}-\inf_{|u|\leq u_{\rm max}}\left\{L_gL_fh_{\rm{i},1}u\right\},\label{eq:ICCBF_hi_2_inequ}
\end{align}
which translates to the BF constraint
\begin{align}\label{eq:bi_2_def}
     b_{\mathrm{i},2}(x):=L_f^2h_{\rm{i},1}&+\left(c_{\rm{i},1}+c_{\rm{i},2} \right)L_fh_{\rm{i},1}\\
     &+c_{\rm{i},1}c_{\rm{i},2}h_{\rm{i},1}+\inf_{|u|\leq u_{\rm max}}\left\{L_gL_fh_{\rm{i},1}u\right\}\geq 0;\nonumber
\end{align}
the manipulation~\eqref{eq:bi_2_def} effectively treats the control term as a disturbance and adds a margin of safety to the dynamics governing $h_{\rm{i},2}$ equal to the disturbance's upper bound. Notice that the relative degree of the ICCBF is no longer two, as is the case for the CBFs in Sections~\ref{sub:ESf} and~\ref{sub:PTSf}; indeed, the ICCBF methodology iterates backstepping \emph{at least once more} by enforcing (similar to~\eqref{eq:hi_2_inequality}) $\frac{d}{dt}b_{\mathrm{i},2}+c_{\mathrm{i},3}b_{\mathrm{i},2}\geq 0$ for $c_{\mathrm{i},3}>0$, which is equivalent to the CBF constraint
\begin{align}
    L_fb_{\mathrm{i},2}(x)+L_gb_{\mathrm{i},2}(x)u+c_{\mathrm{i},3}b_{\mathrm{i},2}(x)\geq 0.
\end{align}
We say that $b_{\mathrm{i},2}$ is an ICCBF if
\begin{align}\label{eq:ICCBF_test_n=2}
    L_fb_{\mathrm{i},2}(x)+\sup_{|u|\leq u_{\rm max}}\left\{L_gb_{\mathrm{i},2}(x)u\right\}+c_{\mathrm{i},3}b_{\mathrm{i},2}(x)\geq 0
\end{align}
holds \emph{only} only on the set $x\in\mathcal{S}_{\rm i}\cap\{b_{\mathrm{i},2}(x)\geq 0\}$ (see~\cite[Def. 4]{agrawal2021safe} for details).

As for standard CBFs, validating~\eqref{eq:ICCBF_test_n=2} is difficult in practice.
However, minimizing the left-hand side of \eqref{eq:ICCBF_test_n=2} is a test to invalidate candidate ICCBFs; in these cases, the authors of~\cite{agrawal2021safe} propose to iterate the backstepping procedure $M\in\mathbb{N}$ times to further restrict the safe set (yet existence of an $M$ guaranteeing~\eqref{eq:ICCBF_test_n=2} is an open problem).

For our LK and OA problems for highway-driving vehicles, we combine the methodologies behind ICCBFs and PTSf (which we call PT-ICCBFs) to investigate performance and safety through simulation studies. While we do not provide a theoretical study of this combination of safety filter designs due to a lack of space, we remark that the interpretation of the control term in~\eqref{eq:ICCBF_hi_2_inequ} as a worst-case disturbance casts the problem into one which was studied in~\cite{steeves2022prescribed} using time-varying backstepping, where positive results were reported.

\section{Low-level MPC}\label{sec:MPC}
The low-level MPC ensures LK via trajectory tracking and runs at a lower sampling rate than the safety-critical control. Hence, we discretize the system~\eqref{eq:ct_sys} and design the control law
\begin{align}\label{eq:control_law}
    u(t_k)=u_s(t_k)+v(t_k),
\end{align}
where $u_s(t_k)$ is sampled from \eqref{eq:us}. The evolution of the discrete-time error signal $e_x(t_{k})=x(t_{k})-x_s(t_k)$ is then described by
\begin{align*}
    e_x(t_{k+1})=A_{\rm d} \,e_x(t_k)+B_{\rm d}\,v(t_k)-w(t_k),
\end{align*}
where subscript ${\rm d}$ denotes the matrices related to the discretized dynamics of \eqref{eq:ct_sys}, using a zero-order-hold, and $w$ denotes the system's deviation from the steady state~\eqref{eq:xs} due to a change in desired yaw rate over one time step. It is defined as
\begin{align}\label{eq:w}
    w(t_k)=A_{\rm d}\Delta x_{\rm s}(t_k)+B_{\rm d}\Delta u_{\rm s}(t_k)+G_{\rm d}\Delta\dot\psi_{\rm ref}(t_{k}),
\end{align}
with $\Delta x_{\rm s}(t_k)=x_{\rm s}(t_{k+1})-x_{\rm s}(t_{k})$, $\Delta u_{\rm s}(t_k)=u_{\rm s}(t_{k+1})-u_{\rm s}(t_{k})$, $\Delta \dot\psi_{\rm ref}(t_k)=\dot\psi_{\rm ref}(t_{k+1})-\dot\psi_{\rm ref}(t_{k})$.
Let the MPC-related cost function be
{
\begin{align*}
    J_N&(e_x(t_k),\mathbf{v}(t_k),\dot{\bm{\psi}}_{\rm ref}(t_k))=|e_x(t_{N|k})|^2_P\\
    &\qquad\qquad \qquad+\sum_{i=0}^{N-1}|e_x(t_{i|k})|^2_Q+|v(t_{i|k})|^2_R,\\
     & \text{s.t. } e_x(t_{0|k})=e(t_k),\\
    & \quad e_x(t_{i+1|k})=A_{\rm d} \,e_x(t_{i|k})+B_{\rm d}\,v(t_{i|k})-w(t_k),
\end{align*}
}
where $\mathbf{v}(t_k)\doteq \{v(t_{0|k}),\dots,v(t_{N-1|k})\}$ and $\dot{\bm{\psi}}_{\rm ref}(t_k)\doteq\{\dot\psi_{\rm ref}(t_k),\dots,\dot\psi_{\rm ref}(t_{N-1})\}$. The matrices $Q$ and $R$ are positive definite, $|x|_Q\doteq \sqrt{x^\top Qx}$, and for $i\in\{0,\dots,N-1\}$ we define the set of admissible control inputs by
{
\begin{align*}
    \V_{t_k}\doteq \{\mathbf{v}(t_k):&\; |u_{\rm s}(t_{k+i})+v(t_{i|k})|\leq u_{\rm max}\}.
\end{align*}
}
At every time instance, the MPC solves the optimization problem $\mathcal{P}(t_k)$:
\begin{align}\label{eq:MPC}
    \min_{\mathbf{v}(t_k)\in\V_{t_k}} J_N(e_x(t_k),\mathbf{v}(t_k),\dot{\bm{\psi}}_{\rm ref}(t_k))
     \text{ s.t. } e(t_{N|k})\in\E,\nonumber
\end{align}
with $\E\in\R^{n_x}$ as the terminal constraint set on the state characterized later.
\begin{rem}
Since $e_1$ is the lateral error \emph{along the local vehicle axis} (see Figure~\ref{fig:schematic}), state constraints would result in a non-convex nonlinear program; hence, we delegate this responsibility to the safety filter.
\end{rem}
For obtaining recursive feasibility and convergence, we make the following standard assumptions.
\begin{ass}\label{ass:controllability}
$(A_{\rm d},B_{\rm d})$ is controllable.
\end{ass}

\begin{ass}[Bounded admissible reference]\label{ass:bounded-yaw-rate}
There exist positive $c_{\psi}<u_{\rm max}/\bar u_{\rm s}$ and $c_{\Delta\psi}$ such that for all $t_k$,
\begin{align*}
    |\dot\psi_{\rm ref}(t_k)|\leq c_{\psi},\quad 
    |\Delta\dot\psi_{\rm ref}(t_k)|\leq c_{\Delta\psi}.
\end{align*}

\end{ass}
Assumption~\ref{ass:bounded-yaw-rate} ensures that the steady-state tuple $(x_{\rm s}(\dot\psi_{\rm ref}),u_{\rm s}(\dot\psi_{\rm ref}))$ is bounded and varies sufficiently slow. In particular, the upper bound on $c_{\psi}$ implies that the reference yaw rate is such that the corresponding steady-state input, $u_{\rm s}(t_k)=\dot\psi_{\rm ref}(t_k)\bar u_{\rm s}$, is in the interior of the constraint set.
\begin{lem}\label{lem:sets:bounded-w_admissible-v}
Assumption \ref{ass:bounded-yaw-rate} implies that in \eqref{eq:w} $w(t_k)\in\W$ for all $t_k$, where the polytope
\begin{align*}
    \W=\left\{w\in\R^{n_x}:\begin{bmatrix}
    I\\-I
    \end{bmatrix}w\leq c_{\Delta\psi} \begin{bmatrix}
    I\\I
    \end{bmatrix}b\right\},
\end{align*}
with $b=A_{\rm d}\bar x_{\rm s}+B_{\rm d}\bar u_{\rm s}+G_{\rm d}$. Furthermore, the polytopic set of admissible control inputs $v(t_k)$ from \eqref{eq:control_law}, 
\begin{align*}
    \bar\V\doteq\left\{v\in\R:\begin{bmatrix}
    1\\-1
    \end{bmatrix}v\leq \begin{bmatrix}
        u_{\rm max}-c_\psi\bar u_{\rm s}\\
        u_{\rm max}+c_\psi\bar u_{\rm s}
    \end{bmatrix}\right\},
\end{align*}
is non-empty.
\end{lem}
Note that $\bar \V$ describes the set of controls $v(t_k)$ that are admissible at all times, i.e.,  $\bar \V\subset \V_{t_k}$ for all $t_k$. On the other hand, $\W$ characterizes the maximum plant deflection related to a change in desired yaw rate. In this fashion, they represent the worst case scenario for the tracking controller and hence the basis for the so-called \emph{maximal invariant constraint admissible (MICA)} set:
\begin{align*}
    \E&=\{e\in\R^{n_x}: (A_{\rm d}-B_{\rm d}K) -w\in\E, Ke\in\bar\V, w\in\W\},
\end{align*}
where $K$ is chosen such that $(A_{\rm d}-B_{\rm d}K)$ is Hurwitz.The MICA set can be computed using, e.g., \cite{borrelli_bemporad_morari_2017,83532}.
\begin{ass}\label{ass:nonempty_set}
The bound $c_{\Delta\psi}$ in Assumption \ref{ass:bounded-yaw-rate} renders $\E\neq \emptyset$.
\end{ass}
This assumption requires the yaw rate to vary sufficiently slow. Note that
Assumptions~\ref{ass:bounded-yaw-rate} and~\ref{ass:nonempty_set} ensure that there exists a $c_{\Delta\psi}>0$ such that $\E\neq \emptyset$ is indeed satisfied.
\begin{prop}[Recursive feasibility]\label{lem:asymptotic-convergence}
Suppose Assumptions~\ref{ass:controllability}--\ref{ass:nonempty_set} hold. If {problem $\mathcal{P}(t_0)$} is feasible, then {$P(t_k)$} is feasible for all $t_k$.
\end{prop}
The proof is standard for $\E$ being MICA, see e.g. \cite{rawlings2009model,borrelli_bemporad_morari_2017,kouvaritakis2015model}.
Recursive feasibility ensures that the MPC generates a control input at all time instances $t_k$ despite a time-varying road curvature and hence desired yaw rate.
\begin{rem}[Stability]\label{rem:stability}
The notion of stability usually assumes a fixed steady state. For any fixed $\overline{\dot \psi}_{\rm ref}$ satisfying Assumption \ref{ass:bounded-yaw-rate} the MPC renders $x_{\rm s}(\overline{\dot \psi}_{\rm ref})$ exponentially stable if, e.g., $P$ of the terminal cost satisfies the related Ricatti equation.
We refer to \cite{MAYNE2000789} for the proof and an overview of methods achieving stability, e.g., scaling the terminal cost and dropping the terminal constraint \cite{limon2006mpc} for computational improvement.
\end{rem}

\section{Simulations}\label{sec:sim}
In our simulation studies,
we use typical highway conditions in the U.S.\footnote{minimum road radius of 1800m, lane width of 3.7m} \cite{highway_US}, and assume a constant longitudinal vehicle velocity of $v_{\rm l}=20\mathrm{m/s}$ while being controlled by the MPC- and CBF-based safety filter in~\eqref{eq:QP_CS_lemma} or~\eqref{eq:QP_CS_lemma_PTSf} on a single-lane road (with $e_{\rm v}=0$ in~\eqref{eq:h_l_def}). The desired path (the center line) is reachable given the vehicle dynamics and steering constraints. The related reference yaw rate and acceleration are provided as time-dependent discrete points and interpolated for the continuous-time safety filter. 
For feasibility of the MPC\footnote{Horizon $N=30$; MPC implemented as QP}, instead of using slack variables as in Remark \ref{rem:stability}, we choose a terminal cost sufficiently large to ensure that the terminal set $\E$ is reached. Due to the high velocity, in some simulations, we impose $u_{\rm max}=5^\circ$ and saturate the controls accordingly. The dynamics from \eqref{eq:ct_sys} are used as a plant model for the MPC, which is applied at a frequency of $20$Hz; the safety filter is computed continuously. The passing time $t_{\rm obs}+T$ is approximated using~\eqref{eq:passing_time_approx}, or more precisely, using the longitudinal velocity and the relative distance along the desired path to the orthogonal projection of the obstacle onto the path. The car width is encoded within the lane width, $w_{\rm l}$, and obstacle radius $r_{\rm obs}$.
We present two simulation studies:
\begin{enumerate*}[label=\Alph*.]
    \item comparing ESf and PTSf (while using ESf for LK) during early obstacle detection scenarios, and;
    \item comparing input-constrained PTSf and PT-ICCBF (while using ICCBF for LK) during late obstacle detection scenarios.
\end{enumerate*}
\subsection{ESf and PTSf during early obstacle detection}
We assume the obstacle is detected $40$m ahead and use the ESf design in Section~\ref{sub:ESf} for LK. We compare the closed-loop performance of MPC with ESf OA to that of MPC with PTSf OA. In Figure~\ref{fig:TI_PI_xy}, we observe that both designs successfully avoid the obstacle while staying on the road, even though the desired trajectory would lead to unsafe operation. For the ESf OA design, we select $c_{{\rm i},i}=15$ for $i=\{1,2\}$ satisfying~\eqref{eq:ESf_gain_c1}, which allows the vehicle to approach the obstacle very closely {(and seemingly match the performance of PTSf)}; this performance is innate to the PTSf OA design {and is desirable because it allows less intervention with the MPC controller, which tracks the desired path well (see Section~\ref{sub:perf_safety_tradeoff} for a discussion on gain tuning).}


\newsavebox{\myboxObst}
\savebox{\myboxObst}{%
\setlength\fheight{1.0cm} 
\setlength\fwidth{.4\columnwidth}
%
%
\definecolor{mycolor1}{rgb}{0.00000,0.44700,0.74100}%
\definecolor{mycolor2}{rgb}{0.85000,0.32500,0.09800}%
\definecolor{mycolor3}{rgb}{0.63500,0.07800,0.18400}%
\definecolor{mycolor4}{rgb}{0.46600,0.67400,0.18800}%
\definecolor{mygray}{rgb}{0.3,0.3,0.3}%
\definecolor{mypurple}{rgb}{0.4940 0.1840 0.5560}
\begin{tikzpicture}

\begin{axis}[%
width=0.951\fwidth,
height=\fheight,
at={(0\fwidth,0\fheight)},
scale only axis,
xmin=96.9036542874301,
xmax=103.7611005714,
xlabel style={font=\color{white!15!black}},
ymin=46.3098381899903,
ymax=50.7852820036336,
ylabel style={font=\color{white!15!black}},
ylabel near ticks,
yticklabel pos=right,
axis background/.style={fill=white},
nodes={scale=0.7, transform shape}
]
\addplot [color=black, line width=1pt]
  table[row sep=crcr]{%
96.2179096590332	49.1652252374138\\
97.6693728456133	49.8916052180454\\
100.327430167957	51.232826384998\\
};
\%addlegendentry{Shoulder}

\addplot [color=black, line width=1pt, forget plot]
  table[row sep=crcr]{%
97.8843034750951	45.862293808626\\
100.670453457417	47.2594929038543\\
104.446845199797	49.1689269935811\\
};
\%addlegendentry{Obstacle}

\draw[fill=mycolor4] (axis cs:101,48) circle[radius=1];
\addplot [color=mycolor4, line width=.5pt]
  table[row sep=crcr]{%
101 48\\
101.1 48.1\\
};
\addplot [color=mycolor1, line width=1.0pt]
  table[row sep=crcr]{%
96.2179096590332	47.0961910002652\\
97.9933792945243	47.9840767562194\\
103.21706677066	50.6176218738301\\
104.42753411909	51.232826384998\\
};
\%addlegendentry{Desired path (center line)}

\addplot [color=mycolor2, line width=1pt]
  table[row sep=crcr]{%
96.2179096590332	47.6468828237259\\
97.5407618334606	48.3228833388616\\
101.246945814896	50.1951697643247\\
103.325134382265	51.232826384998\\
};
\%addlegendentry{Travelled path (ESf)}

\addplot [color=gray, line width=0.5pt]
  table[row sep=crcr]{%
96.2179096590332	46.6923342456173\\
98.1486803255925	47.6782842985533\\
101.852004283626	49.5478602607703\\
104.446845199797	50.8419785175979\\
};
\%addlegendentry{Car contour (ESf)}

\addplot [color=gray, line width=.5pt, forget plot]
  table[row sep=crcr]{%
96.2179096590332	48.6014338316493\\
97.1556567313211	49.0806392760248\\
100.865218149009	50.9546327253017\\
};
\addplot [color=mycolor3, dashed, line width=1.0pt]
  table[row sep=crcr]{%
96.4483262109449	47.7669687628638\\
100.266583172945	49.7019552878456\\
103.322165782015	51.232826384998\\
};
\%addlegendentry{Travelled path (PTSf)}

\addplot [color=mypurple, dashed, line width=0.7pt]
  table[row sep=crcr]{%
96.8345927531823	47.0098042111658\\
100.64918144546	48.9429305439352\\
104.198878025015	50.7213558246018\\
};
\%addlegendentry{Car contour (PTSf)}

\addplot [color=mypurple, dashed, line width=0.7pt, forget plot]
  table[row sep=crcr]{%
96.2179096590332	48.6031138083798\\
99.8839849004295	50.4609800317559\\
101.424571301705	51.232826384998\\
};
\end{axis}

\begin{axis}[%
width=1.227\fwidth,
height=1.227\fheight,
at={(-0.16\fwidth,-0.135\fheight)},
scale only axis,
xmin=0,
xmax=1,
ymin=0,
ymax=1,
axis line style={draw=none},
ticks=none,
axis x line*=bottom,
axis y line*=left
]
\end{axis}
\end{tikzpicture}%
}
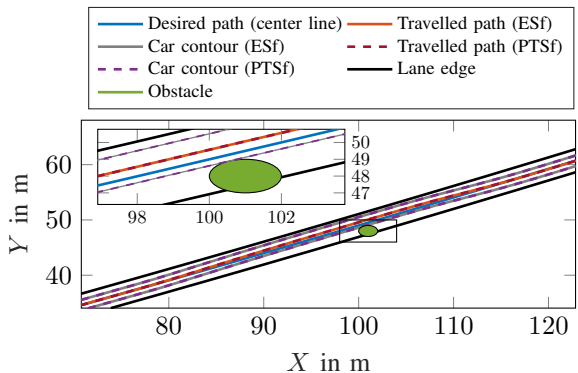
\begin{figure}[t]
    \centering
    \setlength\fheight{2.5cm} 
    \setlength\fwidth{.8\columnwidth}
%
%
\definecolor{mycolor1}{rgb}{0.00000,0.44700,0.74100}%
\definecolor{mycolor2}{rgb}{0.85000,0.32500,0.09800}%
\definecolor{mycolor3}{rgb}{0.63500,0.07800,0.18400}%
\definecolor{mycolor4}{rgb}{0.46600,0.67400,0.18800}%
\definecolor{mygray}{rgb}{0.3,0.3,0.3}%
\definecolor{mypurple}{rgb}{0.4940 0.1840 0.5560}
\begin{tikzpicture}

\begin{axis}[%
width=0.951\fwidth,
height=\fheight,
at={(0\fwidth,0\fheight)},
scale only axis,
xmin=70.7792690437257,
xmax=122.85300176262,
xlabel style={font=\color{white!15!black}},
xlabel={$X$ in m},
ymin=34.0535137727103,
ymax=68.0389152325643,
ylabel style={font=\color{white!15!black}},
ylabel={$Y$ in m},
axis background/.style={fill=white},
legend style={legend cell align=left, align=left, draw=white!15!black,at={(axis cs:71.5,70)},anchor=south west,nodes={scale=0.7, transform shape}},
legend columns=2
transpose legend
]
\addplot [color=mycolor1, line width=1.0pt]
  table[row sep=crcr]{%
70.6672801413256	34.5503457478738\\
76.8114922526903	37.5217496002105\\
82.114195521343	40.1085058373488\\
87.4980510985366	42.7569141738172\\
92.7611603959827	45.3675224997088\\
97.9933792945243	47.9840767562194\\
103.21706677066	50.6176218738301\\
108.432167651113	53.2681300366581\\
113.683089100797	55.9584448194534\\
118.957364078028	58.682654954233\\
122.873045924863	60.7194217507626\\
};
\addlegendentry{Desired path (center line)}

\addplot [color=mycolor2, line width=1.0pt]
  table[row sep=crcr]{%
70.6673198995091	34.550263261051\\
76.811601865518	37.5215241510939\\
78.4329753631625	38.3179209235757\\
79.4725108947733	38.8511476156632\\
80.9952068031756	39.6576126425699\\
87.5648537717524	43.1668028779217\\
90.5999789505774	44.7561707912758\\
93.8432740125132	46.4334026243968\\
97.5407618334606	48.3228833388616\\
101.246945814896	50.1951697643247\\
104.514649771009	51.826761188559\\
106.270162617022	52.6841595975796\\
107.607491005094	53.3175194090363\\
108.930273148179	53.9229841202999\\
111.168551645347	54.9213899617085\\
113.215708813357	55.8421001397673\\
114.52842645538	56.4512172934445\\
115.857526642451	57.0877592059737\\
117.291511607902	57.7971270075032\\
118.972429957482	58.6538053788127\\
121.024030061518	59.7239646337918\\
122.881326488311	60.7038773793965\\
};
\addlegendentry{Travelled path (ESf)}

\addplot [color=gray, line width=1.0pt]
  table[row sep=crcr]{%
71.4869149703336	34.0016713956309\\
77.4082382475676	36.8664025549822\\
78.8085217690329	37.5553824674715\\
79.858109440499	38.093642657571\\
81.3973990870476	38.9087859825251\\
87.960513674634	42.4145042395793\\
90.991844072858	44.0018885324761\\
94.4544113422739	45.7918019924213\\
98.1486803255925	47.6782842985533\\
101.852004283626	49.5478602607703\\
105.116621943222	51.1760145296955\\
106.866777486096	52.0276770844039\\
108.196080139814	52.6535263621981\\
109.7348798716	53.3528471515512\\
114.2239808657	55.3706290860906\\
115.546173958514	55.9936338540484\\
116.826512896116	56.6160754408609\\
118.32641284183	57.3677242028211\\
120.161200009268	58.3128208614481\\
122.640139549276	59.6150217598845\\
123.060031774088	59.8368890896483\\
};
\addlegendentry{Car contour (ESf)}

\addplot [color=gray, line width=1.0pt, forget plot]
  table[row sep=crcr]{%
70.748583723306	35.5329672421814\\
76.6646847975461	38.3951694655873\\
78.0574289572921	39.08045937968\\
79.0869123490475	39.6086525737553\\
80.5930145193036	40.4064393026147\\
87.1691938688708	43.919101516264\\
90.2081138282967	45.5104530500755\\
93.4543713228297	47.1892165551741\\
97.1556567313211	49.0806392760248\\
100.865218149009	50.9546327253017\\
104.136624411274	52.5880737573485\\
105.898329324291	53.4485155371893\\
107.24461001809	54.0861658747609\\
108.578945045224	54.6969793167152\\
110.829194430283	55.7007080486176\\
112.869685664671	56.6184814768521\\
114.394777295857	57.3281480918005\\
115.852540606999	58.0314755009418\\
117.351700181477	58.7794471996538\\
119.021262339074	59.6361486671854\\
121.235030968774	60.7956911590093\\
122.901798059638	61.6762294666636\\
};
\addplot [color=mycolor3, dashed, line width=1.0pt]
  table[row sep=crcr]{%
70.6673198795312	34.5502633025007\\
76.2037837112761	37.2279501085017\\
77.9516650355646	38.0923064084365\\
79.4666654129088	38.8624206508609\\
80.9945856487367	39.6599908298837\\
83.1312282184366	40.8002261378017\\
86.9226395468259	42.8273483541118\\
89.7352357137902	44.3054292532999\\
92.976176889042	45.9869483275656\\
96.4483262109449	47.7669687628638\\
100.266583172945	49.7019552878456\\
103.819694267091	51.4820920980416\\
105.796357419237	52.4550598529342\\
107.334796027562	53.1899763446433\\
108.679081455389	53.8092235068109\\
110.666702808844	54.6981068953352\\
112.943509157109	55.7177329216721\\
114.483322556302	56.4297992841682\\
115.812634735707	57.0657902700478\\
117.246850804596	57.774632424213\\
118.719918339322	58.5236382284285\\
120.758442972398	59.5844415800047\\
123.058117183993	60.7973747462147\\
};
\addlegendentry{Travelled path (PTSf)}

\addplot [color=mypurple, dashed, line width=1pt]
  table[row sep=crcr]{%
71.4869149541461	34.0016714568687\\
76.5757873342997	36.463677052286\\
78.3296767670768	37.3309870727835\\
79.8539342895779	38.105768275484\\
81.3919264672963	38.9085786434597\\
83.5372211361297	40.0534532467679\\
87.322174210086	42.0771003598768\\
90.1277352392683	43.5514769189316\\
93.3656748722399	45.2314410006092\\
96.8345927531823	47.0098042111658\\
100.64918144546	48.9429305439352\\
104.198878025015	50.7213558246018\\
106.170405191884	51.6917851554777\\
107.699795542226	52.4223336189596\\
109.032589472111	53.0362215132082\\
111.007106466726	53.9192453224605\\
113.287749077361	54.940559274856\\
114.840132383638	55.658315812462\\
116.182505977417	56.3004829693142\\
117.629555536268	57.0156613520664\\
119.112185657636	57.7695650549346\\
121.157212711639	58.8337867358416\\
123.015834480188	59.8135445411631\\
};
\addlegendentry{Car contour (PTSf)}

\addplot [color=mypurple, dashed, line width=1pt, forget plot]
  table[row sep=crcr]{%
70.7485836854758	35.5329672929838\\
76.0558957392884	38.1018790576207\\
77.7958365185415	38.965253232511\\
79.299894772911	39.7329387831785\\
81.0141419868964	40.6321374221729\\
83.5154163279889	41.9716301050879\\
86.5231048835658	43.5775963483468\\
89.3427361883121	45.0593815876683\\
92.5866789058442	46.742455654522\\
96.0620596687076	48.5241333145618\\
99.8839849004295	50.4609800317559\\
103.440510509168	52.2428283714814\\
105.422309646589	53.2183345503906\\
106.969796512898	53.957619070327\\
108.325573438668	54.5822255004135\\
110.326299150963	55.4769684682099\\
112.599269236858	56.4949065684882\\
114.126512728965	57.2012827558745\\
115.442763493998	57.8310975707814\\
116.864146072923	58.5336034963596\\
118.534762672275	59.3844162768684\\
120.580694407142	60.4511949827942\\
122.857568373636	61.6528548264761\\
};
\addplot [color=black, line width=1pt]
  table[row sep=crcr]{%
70.7417409528411	36.6399530792078\\
76.4968226062047	39.4257213739344\\
81.7484390499102	41.9894501179813\\
87.1260268661462	44.6366288185269\\
92.3606509204652	47.2348727808601\\
97.6693728456133	49.8916052180454\\
102.98265699806	52.5726193223654\\
108.191405388983	55.2223553314891\\
113.435922631482	57.9118704547649\\
118.701812498213	60.6342456423149\\
123.057373869435	62.9026582425234\\
};
\addlegendentry{Lane edge}

\addplot [color=black, line width=1pt, forget plot]
  table[row sep=crcr]{%
73.8817010080095	34.0469546443884\\
79.420653725779	36.7337541175974\\
84.8370717877554	39.383667171503\\
90.0675152030848	41.9639663567439\\
95.3331989024068	44.5829653362148\\
100.670453457417	47.2594929038543\\
105.972671949322	49.9404215336906\\
111.189959175526	52.5998889748328\\
116.443056152387	55.2992467589444\\
121.598466185979	57.9696219111062\\
122.876615923974	58.6349352515407\\
};

\draw[fill=mycolor4] (axis cs:101,48) circle[radius=1];
\addplot [color=mycolor4, line width=1pt]
  table[row sep=crcr]{%
101 48\\
101.1 48.1\\
};
\addlegendentry{Obstacle}

\node at (axis cs: 84,58) {\usebox{\myboxObst}};
\draw (axis cs: 98,46) rectangle (axis cs: 104,50); 
\end{axis}

\begin{axis}[%
width=1.227\fwidth,
height=1.227\fheight,
at={(-0.16\fwidth,-0.135\fheight)},
scale only axis,
xmin=0,
xmax=1,
ymin=0,
ymax=1,
axis line style={draw=none},
ticks=none,
axis x line*=bottom,
axis y line*=left
]
\end{axis}
\end{tikzpicture}%
  \caption{The OA designs override the MPC control to enforce safety. The ESf filter is tuned to approximately match the performance of the PTSf one. Both designs pass the obstacle at the boundary of the safe set.}
    \label{fig:TI_PI_xy}
\end{figure}

\begin{figure}[t]
\vspace{-5pt}
    \centering
    \setlength\fheight{2.5cm} 
    \setlength\fwidth{.8\columnwidth}
    \input{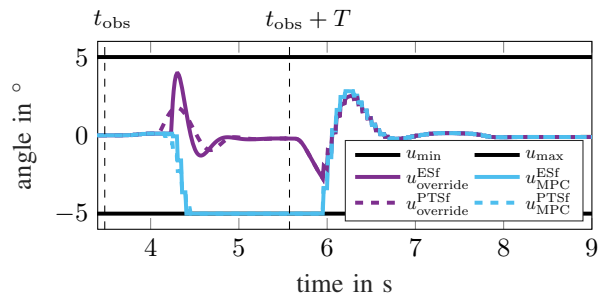}
  \caption{For similar closed-loop vehicle trajectories, the PTSf filter overrides the MPC control less aggressively than the ESf filter. The discrete-time MPC sends the minimum steering command as the vehicle deviates from the desired trajectory.}
    \label{fig:TI_PI_u}
    \vspace{-10pt}
\end{figure}

{We now investigate the control effort required for this OA task}. It is clear from Figure~\ref{fig:TI_PI_u} that the nominal MPC control input, which seeks to track the centerline, is overwritten by the safety filters to avoid the obstacle. The filters mainly differ early on when safety constraints become active. We observe a large {control input generated by the} ESf OA design {which is undesirable (see ~\cite{abel2022} for a further discussion)}.
We can alleviate this by {lowering} the gains $c_{{\rm i},i}$ at the cost of increased conservatism, which {can become} problematic when the lane width is limited, {potentially violating the control sharing property.}
The PTSf OA controller has a significantly smaller peak but allows the vehicle to approach the barrier equally to the ESf OA design.
After passing the obstacle, control authority is gradually ceded to the MPC using a smooth step function similar to~\eqref{eq:Phi_def}; see~\cite{abel2022} for details.

Throughout the maneuver, the control sharing property among the LK and OA controllers was verified. The control inputs are within the input constraint set for all time, {but are not designed to do so.}

\subsection{PTSf and PT-ICCBF during late obstacle detection}
Now suppose that the obstacle is detected when only $15$m ahead of the vehicle. We saturate the inputs of the MPC with ESf for LK and PTSf for OA control designs to $|u|\leq u_{\rm max}$ and compare the results to MPC with the input-constrained equivalent designs following the methodology in Section~\ref{sub:ICCBF}. Figure~\ref{fig:IC_xy} illustrates the non-constrained designs steering the vehicle away from the obstacle, but due to the input saturation, violate safety since the vehicle contour intersects with the obstacle. Furthermore, once the obstacle is passed, this design induces oscillations which culminate in the vehicle driving off the road. In comparison, the ICCBF LK with PT-ICCBF OA design successfully avoids the obstacle while steering the vehicle to the barrier, and additionally, does not cause the vehicle to exit the road after passing the obstacle.
\newsavebox{\myboxObstIC}
\savebox{\myboxObstIC}{%
\setlength\fheight{1.5cm} 
\setlength\fwidth{.4\columnwidth}
%
%
\definecolor{mycolor1}{rgb}{0.00000,0.44700,0.74100}%
\definecolor{mycolor2}{rgb}{0.85000,0.32500,0.09800}%
\definecolor{mycolor3}{rgb}{0.63500,0.07800,0.18400}%
\definecolor{mycolor4}{rgb}{0.46600,0.67400,0.18800}%
\definecolor{mygray}{rgb}{0.3,0.3,0.3}%
\definecolor{mypurple}{rgb}{0.4940 0.1840 0.5560}
\begin{tikzpicture}

\begin{axis}[%
width=0.951\fwidth,
height=\fheight,
at={(0\fwidth,0\fheight)},
scale only axis,
xmin=94.6532061436866+2,
xmax=106.358738836874-2,
xlabel style={font=\color{white!15!black}},
xlabel near ticks,
ymin=45.1793766135105+1.5,
ymax=52.2026962294227-1.5,
ylabel style={font=\color{white!15!black}},
axis background/.style={fill=white},
nodes={scale=0.6, transform shape},
ylabel near ticks,
yticklabel pos=right
]

\addplot [color=mycolor1, line width=1.0pt]
  table[row sep=crcr]{%
94.5509095236101	46.2601577844219\\
96.8825543786426	47.4267915841176\\
99.3693375307303	48.6757073951682\\
101.860185602677	49.9314963870668\\
104.265235197306	51.1486284089967\\
106.427379292246	52.2466981365814\\
};

\addplot [color=mycolor2, line width=1.0pt]
  table[row sep=crcr]{%
94.5056240940112	46.7914828039064\\
94.9491893812004	47.0327751461868\\
95.3731258267587	47.2587503592406\\
95.8220307039373	47.4927699615136\\
96.1896106056977	47.6801659278318\\
96.6432846028973	47.9060058848671\\
97.0699699365994	48.1127674481722\\
97.5053466715953	48.317985571196\\
97.9640093205554	48.530624468694\\
98.3959672495256	48.7342150726634\\
98.9566235305264	49.0023348651241\\
99.6231220352783	49.3257511326471\\
100.277690379491	49.6487259868709\\
100.73913212701	49.8803398687283\\
101.173965283236	50.1022160029703\\
101.648827362767	50.3491972445184\\
102.085408569224	50.5811799482296\\
102.493887093738	50.8029757821673\\
102.925277608573	51.0426099657596\\
103.33236891665	51.2740878677347\\
103.757950799013	51.5217377356445\\
104.159496461018	51.760672621278\\
104.579391838984	52.0158529554713\\
104.975824261237	52.2615503724698\\
};

\addplot [color=mycolor3, dashed, line width=1.0pt]
  table[row sep=crcr]{%
94.4770253623849	46.5805062004871\\
95.0978102500198	46.9454315740946\\
95.6804699700789	47.282734086454\\
96.1166584080806	47.5310182497714\\
96.53353677948	47.7646600379533\\
96.9750061892251	48.0078876807298\\
97.3969295800104	48.2359923797271\\
97.8437096566525	48.472561371211\\
98.2706337751759	48.6935542215163\\
99.0395474358688	49.0826767327008\\
101.006747839192	50.0744687257121\\
103.124425351405	51.1407764024098\\
104.18286936309	51.6797999431441\\
105.320436730784	52.264181425962\\
};

\addplot [color=mypurple, dashed, line width=1pt]
  table[row sep=crcr]{%
94.6178117985708	45.6761598583773\\
95.3273062263787	46.0950686606845\\
95.9044621384785	46.4311781725193\\
96.3345331699591	46.6776793763974\\
96.7443730128636	46.9090915343865\\
97.1774215337127	47.1496560373327\\
97.5906018674094	47.3751009814314\\
98.0275888261954	47.6088643237081\\
98.4447950124742	47.8272706340113\\
98.9429593715617	48.0816408644768\\
101.959269421385	49.6018836400675\\
103.279305145913	50.2665110390843\\
104.344826734929	50.8079722967693\\
105.488628996559	51.394698200081\\
106.378592715081	51.8534139184232\\
};

\addplot [color=mypurple, line width=1pt, dashed, forget plot]
  table[row sep=crcr]{%
94.4374089631494	47.5435447084316\\
95.0223057891363	47.8843815909419\\
95.4619226857156	48.1366348050478\\
95.8832718939744	48.3748694381113\\
96.3304398352678	48.6236377205633\\
96.7585135238717	48.857541195588\\
97.2123742287645	49.1006502199196\\
97.6464817364515	49.3281690027645\\
98.165513413119	49.5934536113807\\
100.790991803152	50.9172235374346\\
102.525083341411	51.7900429454171\\
103.356841290143	52.2123571294965\\
};

\addplot [color=gray, line width=1.0pt]
  table[row sep=crcr]{%
94.5160646697214	45.825778348319\\
94.927872155266	46.0537792219938\\
95.3633343065795	46.2904921942185\\
95.7790604775288	46.5119457935715\\
96.2189134021698	46.7411157033313\\
96.5788824247686	46.9245420459452\\
97.0229877881648	47.1455287378747\\
97.4405492855406	47.3478027783094\\
97.8665317643212	47.5485407526466\\
98.3163954867109	47.757110415879\\
98.743870028599	47.9586741928169\\
99.3034914298648	48.2263305736356\\
99.9741566328566	48.5516227768152\\
100.636864610607	48.8783404524127\\
101.105715283189	49.1134521392619\\
101.548422908266	49.3391422928632\\
102.032597366483	49.5907642651146\\
102.47820669067	49.8273831348981\\
102.895413111237	50.0537916544182\\
103.336204264281	50.2985405579562\\
103.752221830154	50.5350184995273\\
104.187044136663	50.7879946555516\\
104.59707176992	51.0319560446588\\
105.025431220652	51.2922861859185\\
105.429338440434	51.5426450315048\\
105.851450438241	51.8090484569034\\
106.458966970122	52.1989785512869\\
};

\addplot [color=gray, line width=1.0pt, forget plot]
  table[row sep=crcr]{%
94.5350444558213	47.7750580981551\\
94.9671911759885	48.0055549249096\\
95.4251480057047	48.244424219696\\
95.8003387866269	48.4357898097183\\
96.2635814176299	48.6664830318595\\
96.6993905876581	48.877732118035\\
97.1441615788693	49.0874303897454\\
97.8192312445477	49.4015596610462\\
98.500813450564	49.7259429824001\\
99.0504904528418	49.9917224146712\\
99.6994250974701	50.3102469890252\\
100.311650424145	50.6164182969\\
100.799507658205	50.8652897130773\\
101.26505735905	51.1076302239221\\
101.692610447779	51.3349767615612\\
102.092361076239	51.5521599099164\\
102.514350952866	51.7866793735629\\
102.912516003146	52.0131572359421\\
103.328857461363	52.2554808157375\\
};
\addplot [color=black, line width=1pt]
  table[row sep=crcr]{%
94.4631368765691	48.284434393858\\
96.8056546161706	49.4578666886462\\
99.2472855540783	50.6854970812905\\
101.645695582501	51.8959277749515\\
102.291964009918	52.2228562438834\\
};

\addplot [color=black, line width=1pt, forget plot]
  table[row sep=crcr]{%
96.4972146127997	45.1648098569083\\
98.9134413443183	46.3759415379624\\
101.352601088691	47.6031733330295\\
103.740431105607	48.8090710834007\\
106.173496505306	50.0423951219453\\
106.396619949621	50.1557285500646\\
};

\draw[fill=mycolor4] (axis cs:101,48) circle[radius=1];
\addplot [color=mycolor4, line width=.5pt]
  table[row sep=crcr]{%
101 48\\
101.1 48.1\\
};
\end{axis}

\begin{axis}[%
width=1.227\fwidth,
height=1.227\fheight,
at={(-0.16\fwidth,-0.135\fheight)},
scale only axis,
xmin=0,
xmax=1,
ymin=0,
ymax=1,
axis line style={draw=none},
ticks=none,
axis x line*=bottom,
axis y line*=left
]
\end{axis}
\end{tikzpicture}%
}
\newsavebox{\myboxRoadIC}
\savebox{\myboxRoadIC}{%
\setlength\fheight{1.5cm} 
\setlength\fwidth{.4\columnwidth}
%
%
\definecolor{mycolor1}{rgb}{0.00000,0.44700,0.74100}%
\definecolor{mycolor2}{rgb}{0.85000,0.32500,0.09800}%
\definecolor{mycolor3}{rgb}{0.63500,0.07800,0.18400}%
\definecolor{mycolor4}{rgb}{0.46600,0.67400,0.18800}%
\definecolor{mygray}{rgb}{0.3,0.3,0.3}%
\definecolor{mypurple}{rgb}{0.4940 0.1840 0.5560}
\begin{tikzpicture}

\begin{axis}[%
width=0.951\fwidth,
height=\fheight,
at={(0\fwidth,0\fheight)},
scale only axis,
xmin=105.619979830329,
xmax=117.325512523517,
xlabel style={font=\color{white!15!black}},
xlabel near ticks,
xticklabel pos=top,
ymin=53.2620167252328,
ymax=60.285336341145,
ylabel style={font=\color{white!15!black}},
axis background/.style={fill=white},
nodes={scale=0.6, transform shape},
ylabel near ticks,
yticklabel pos=right
]
\addplot [color=mycolor1, line width=1.0pt]
  table[row sep=crcr]{%
108.298554827593	53.1999566184585\\
110.680463273452	54.4173867165618\\
113.049271024033	55.6325536304507\\
115.372183373749	56.8284794674178\\
117.326828122146	57.8381147713901\\
};

\addplot [color=mycolor2, line width=1.0pt]
  table[row sep=crcr]{%
106.406772780751	53.1728766805721\\
106.96493819604	53.5234287570424\\
107.552211319396	53.8875082734818\\
107.983044276946	54.1511730105171\\
108.395529072825	54.4004220208059\\
108.833064859101	54.6609006568039\\
109.251877661633	54.9060079863029\\
109.696016023323	55.1609806366524\\
110.12101123726	55.3998358901933\\
110.571517104096	55.6472487831529\\
111.025137274964	55.8901217622679\\
111.458840138982	56.1163037973688\\
111.918132355337	56.3492925573577\\
112.322028801873	56.5485661441634\\
112.786169947345	56.7710443325111\\
113.171015123902	56.9502233181337\\
113.639220892841	57.1617619879819\\
114.085806324682	57.3569621852404\\
114.557568464315	57.5562431721229\\
115.007127224221	57.7395735127657\\
115.481580966634	57.9261553472922\\
115.9332711042	58.0972528429469\\
116.409092273699	58.2712128938708\\
117.331047972392	58.5997765440089\\
};

\addplot [color=gray, line width=1.0pt]
  table[row sep=crcr]{%
108.017257855277	53.1760086463104\\
108.442534840425	53.4360727223944\\
108.848686889271	53.6812919954327\\
109.278703410685	53.937086950563\\
109.689751684745	54.1774708631009\\
110.125221395966	54.4273030861615\\
110.541625092071	54.6611993184335\\
110.982805794989	54.903379431221\\
111.426884258833	55.1410561036289\\
111.851371606395	55.3623680924899\\
112.300842562709	55.5903242463112\\
112.69606989121	55.7852881715429\\
113.15022681446	56.0029541050596\\
113.526782808023	56.178258710826\\
113.984895691556	56.3852254894622\\
114.421847708129	56.5762085822317\\
114.883425938071	56.7711846349438\\
115.323271018309	56.9505531698975\\
115.787460297883	57.1330993374702\\
116.229362440187	57.3004906947786\\
116.695095818355	57.470774281331\\
117.369964873207	57.7114522677841\\
};

\addplot [color=gray, line width=1.0pt, forget plot]
  table[row sep=crcr]{%
105.520616806995	53.6165465289727\\
106.145337927442	54.0125697808808\\
106.708772696486	54.364861728862\\
107.30476552962	54.7326741680999\\
107.743485771715	54.9997006014541\\
108.164376083374	55.252367183233\\
108.611494098575	55.5164977551661\\
109.039934375685	55.7650093993856\\
109.494609177797	56.0234144592419\\
109.929884989487	56.2653389834028\\
110.3914046052	56.5157520764705\\
110.832850074309	56.7492137050653\\
111.300507553508	56.9899347219941\\
111.711881360677	57.195995963837\\
112.184740288135	57.4262329471203\\
112.576908690253	57.6117964051745\\
113.054135691216	57.8310287926633\\
113.509439011874	58.0334821978898\\
113.99053181508	58.2403273009105\\
114.449100848628	58.4307674305405\\
114.933193611398	58.6247455646073\\
115.394188712126	58.8027804948414\\
115.880325173057	58.9835606717417\\
116.365408225619	59.1583348783208\\
117.505430812993	59.5639297327034\\
};
\addplot [color=mycolor3, dashed, line width=1.0pt]
  table[row sep=crcr]{%
107.039389826146	53.1495639525608\\
107.990757974044	53.6352878388459\\
108.773368943342	54.0295347600351\\
109.448495852386	54.3642335411121\\
110.081477390073	54.6724145459425\\
110.512579000483	54.8787238329706\\
110.967831711649	55.0931070244627\\
111.401721493067	55.2938451775912\\
112.088511432067	55.6056809427588\\
112.753383275593	55.9019603679821\\
113.646681267497	56.2957139831473\\
115.173965280642	56.9683977844687\\
115.995358193677	57.3348748159049\\
116.676351865816	57.6430762622342\\
117.333268623984	57.9449804298886\\
};

\addplot [color=mypurple,dashed, line width=1pt]
  table[row sep=crcr]{%
109.031804722205	53.2083984457877\\
109.606071689605	53.4942911045605\\
110.232496488094	53.8012576123229\\
110.680953779869	54.0173920757166\\
111.107883726135	54.2200188843163\\
111.558275293379	54.4301869748009\\
112.213057869659	54.7294537346239\\
112.869618157824	55.023526009796\\
113.754964654935	55.4147750746312\\
115.736632616131	56.2882823903931\\
116.562605911763	56.6581814991524\\
117.226275604593	56.9599038545293\\
117.455052448075	57.064983659363\\
};

\addplot [color=mypurple, line width=1pt, dashed, forget plot]
  table[row sep=crcr]{%
105.593081525399	53.3610518626543\\
106.869261393054	54.0175194181489\\
107.600227211019	54.3902618272268\\
108.387131365227	54.7867140870456\\
109.067408112031	55.1240178092771\\
109.706346606517	55.4351575532708\\
110.1420862446	55.6437304459079\\
110.602713299848	55.860693205529\\
111.042151739148	56.0640461851642\\
111.73776385715	56.3799393871365\\
112.41040064692	56.6796897016376\\
113.310962330553	57.0766062903047\\
114.839888841541	57.7499941846893\\
115.656956103207	58.1146081197758\\
116.332374768382	58.4203662701728\\
116.982788653419	58.7193600462483\\
117.407800558776	58.9174024181147\\
};
\addplot [color=black, line width=1pt]
  table[row sep=crcr]{%
105.588107648586	53.8953712722114\\
108.041635054333	55.1458680658571\\
110.325831778463	56.3143211722207\\
112.703109686577	57.5347605843019\\
115.076257221835	58.7575370488037\\
117.385685485731	59.951767422504\\
};

\addplot [color=black, line width=1pt, forget plot]
  table[row sep=crcr]{%
112.437256969876	53.2388561970038\\
114.841354700959	54.4738888479265\\
117.132423271909	55.6551023833953\\
117.354767077122	55.7699578352879\\
};

\end{axis}

\begin{axis}[%
width=1.227\fwidth,
height=1.227\fheight,
at={(-0.16\fwidth,-0.135\fheight)},
scale only axis,
xmin=0,
xmax=1,
ymin=0,
ymax=1,
axis line style={draw=none},
ticks=none,
axis x line*=bottom,
axis y line*=left
]
\end{axis}
\end{tikzpicture}%
}
\begin{figure*}[t]
    \centering
    \setlength\fheight{3cm} 
    \setlength\fwidth{.8\textwidth}
    \input{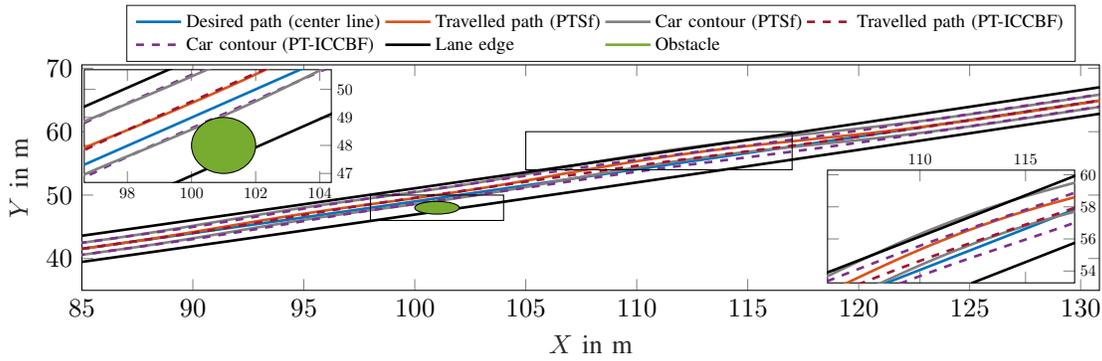}
  \caption{The vehicle leaves the desired trajectory to avoid the obstacle. The ICPTSf renders the closed loop safe. On the contrary, the zooms reveal that the PTSf violates safety (both of the road barrier and the obstacle) due to saturated control inputs.}
    \label{fig:IC_xy}
\end{figure*}
Due to the late obstacle detection, the control inputs for both safety filters in Figure~\ref{fig:IC_u} are of large amplitude and are saturated at their maximum for much of the time. Despite being saturated, the ICCBF for LK with PT-ICCBF for OA controller ensures safe vehicle operation regardless of being saturated to the maximum allowable, whereas this same input saturation renders the controller absent of the input-constrained design consideration unsafe.
\begin{figure}[t]
    \centering
    \setlength\fheight{2.5cm} 
    \setlength\fwidth{.8\columnwidth}
    \input{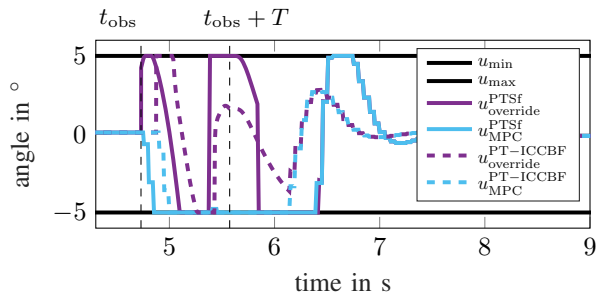}
  \caption{For a similar trajectory the PT safety filter shows a smoother interjection with less jerk. We also observe the discrete-time nature of the MPC, which sends the minimum steering command as the vehicle is off the desired trajectory.}
    \label{fig:IC_u}
\end{figure}

\section{Conclusion and Acknowledgements}
\label{sec:conclusions}
Our multi-layer MPC and CBF-Safety design exploits the advantages (numerical cost, nonlinear-model-based control fidelity, and ability to perform swift interventions) of both control strategies while allowing the encoding of safety prioritization of OA over LK. This prioritization allows us to establish CBF-QP feasibility, but is conservative, as our simulation studies which do not require any prioritization, show. Additional to ESf filter designs, we explore PTSf and input-constrained safety designs, which bring the advantages of retaining safety while balancing performance, and practicality. Our ongoing research aims to provide some theoretical guarantees for the combinations of ESfs, PTSfs and their input-constrained counterparts for simultaneous LK and OA.

The authors thank Imoleayo Abel and Bob Bitmead for fruitful discussions and valuable comments.
\bibliographystyle{ieeetr}
\bibliography{bib_all}

\end{document}